\documentclass[
 amsmath,amssymb,
 aps,
]{revtex4-2}

\usepackage{graphicx}%
\usepackage{dcolumn}%
\usepackage{bm}%
\usepackage{subcaption}
\usepackage{xcolor}
\usepackage{makecell}

\makeatletter
\newcommand{\fmarki}{*}
\newcommand{\fmarkii}{\ensuremath{\mathsection}}
\newcommand{\fmarkiii}{\ensuremath{\mathparagraph}}
\newcommand{\fmarkiv}{\ensuremath{\ddagger}}
\newcommand{\fmarkv}{\ensuremath{\|}}
\newcommand{\fmarkvi}{**}
\newcommand{\fmarkvii}{\ensuremath{\dagger\dagger}}
\newcommand{\fmarkviii}{\ensuremath{\ddagger\ddagger}}
\def\@fnsymbol#1{{%
  \ifcase#1\or
  \fmarki\or \fmarkii\or \fmarkiii\or \fmarkiv\or
  \fmarkv\or \fmarkvi\or \fmarkvii\or \fmarkviii
  \else\@ctrerr
  \fi}}
\makeatother

\newcommand{\uin}{\mathbf{u}_{\mathrm{in}}}
\newcommand{\uout}{\mathbf{u}_{\mathrm{out}}}
\newcommand{\dFR}{d_{\mathrm{FR}}}
\newcommand{\DB}{D_{\mathrm{B}}}

\begin{document}

\preprint{}

\title{What should a linear optical frontend compute? Assessing the role of meta-optics, nonlocality, and coherence in hybrid inference systems}

\author{Nicholas Behrens}
\thanks{These authors contributed equally to this work}

\author{Yandong Li\,$^*$}
 \email{yl2695@cornell.edu}

\author{Francesco Monticone}
 \email{francesco.monticone@cornell.edu}
\affiliation{School of Electrical and Computer Engineering, Cornell University, Ithaca, New York 14853, USA}

\date{\today}

\begin{abstract}
Hybrid inference systems that pair an optical frontend with a digital backend offer a route to offload computation to the physical layer. Yet what the optics should compute, and when this is beneficial, has remained unclear. Here, we show that, for classification tasks, a well-designed optical frontend reshapes the statistics of the sensor intensity readout to improve class separability, quantified by the Bhattacharyya distance. This training-free metric predicts downstream accuracy and reveals that much of the discriminative information resides in inter-pixel correlations. We then identify the roles of coherence and different forms of nonlocality. Because the sensor measures intensity, a linear frontend produces features that are quadratic in the input field; however, only nonlocal, coherent optical systems can exploit the associated information. Such systems can yield significant performance gains, surpassing the best trained linear preprocessor. These results provide physical insights and new design principles for optimal optical--electronic inference systems.
\end{abstract}

\maketitle

Recent advances in AI have led to increasing demands for computational resources and better implementation strategies to accelerate training and inference~\cite{ai_and_compute:2018,scaling_law_DeepMind:2022}. This has renewed interest in optical computing, which aims to exploit the passive, large-scale spatial and spectral parallelism of light propagation and scattering to perform analog operations with very low latency and energy~\cite{McMahon:2023,Li:2024}. An important class of optical computing architectures consists of hybrid two-stage systems, in which an optical frontend performs large-scale preprocessing, and an electronic backend carries out the downstream task, for example classification, or more general-purpose decision-making tasks. In particular, optical frontends leveraging the large number of degrees of freedom offered by metasurfaces, co-optimized with digital backends, have been demonstrated for classification~\cite{Huang2023,Wei:2024}, segmentation~\cite{liu2024extrememeta,choi2026meta}, image reconstruction~\cite{Tseng:2021,lin2021end}, and, at large scale, general-purpose edge vision tasks under natural light~\cite{Peng:2026}.
The promise of such systems is to offload computation from the electronic backend, reducing the number of stored parameters and the energy spent per inference, while exploiting the high throughput, low dissipation, and low latency of free-space optics. Realizing this promise requires a better understanding of \emph{how much}, and \emph{what}, the optics should compute. More broadly, whether, when, and to what extent this strategy is beneficial remains an active area of research. 

\begin{figure}[htbp]
    \centering
    \includegraphics[width=0.7\textwidth]{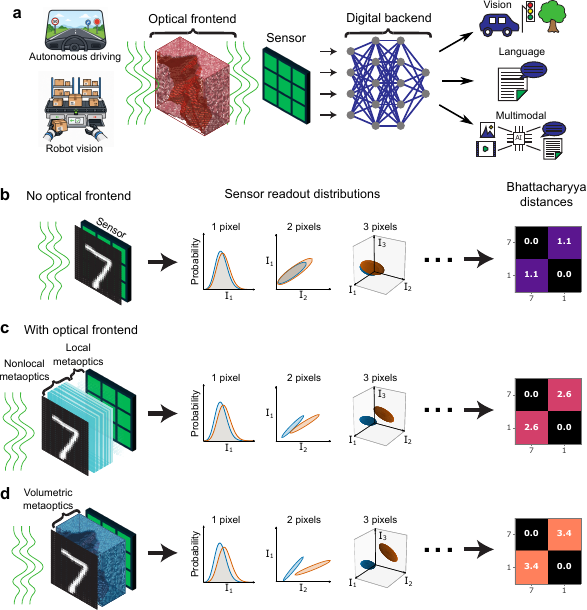}
    \caption{\label{fig:Schematic}
    \textbf{Hybrid end-to-end-optimized optical--electronic inference systems.}
    (a) Illustration of a hybrid optical--electronic inference pipeline. A coherent or incoherent field carrying the input propagates through an optical frontend and is then measured by a sensor array, which acts as a bottleneck between the optical frontend and the electronic backend. The digital backend is a neural network that supports downstream applications, such as vision, language, and multimodal tasks. 
    (b--d) Frontends considered here: (b) a sensor-only configuration (baseline), with an ideal fixed lens (not shown) imaging the input onto the sensor; (c) a frontend combining a \emph{local} phase mask (e.g., a local metasurface) with a \emph{nonlocal} $k$-space-filtering component; and (d) a general volumetric operator (e.g., a 3D volumetric metamaterial). The frontend reshapes the distributions of the intensity readouts associated with different classes. The single-pixel and multi-pixel (joint) distributions of two example classes are illustrated in the middle panels. The distribution separation is quantified by a statistical distance, the Bhattacharyya distance.
    }
\end{figure}

Importantly, the hybrid nature of these architectures calls for new design principles and strategies to bridge the optical-analog and electronic-digital domains effectively and efficiently, for the given task and available resources. A common design approach is to consider a low-dimensional ``bottleneck'' between the two stages (Fig.~\ref{fig:Schematic}(a)): a small sensor array limits both the analog-to-digital conversion and the input dimension---and hence the size---of the electronic backend. Constraining the throughput in this way, so as to achieve compression while retaining task-relevant information, is reminiscent of the \emph{information bottleneck} principle~\cite{information_bottleneck,shwartzziv:2017}, which was introduced to understand the tradeoffs between compression and accuracy in signal processing and, more recently, to address information-theoretic questions in deep learning.

These considerations raise several fundamental questions. What specific operation should an optical frontend perform to encode the input so that task-relevant information survives the bottleneck? Equivalently, among all possible optical responses allowed by physics, what should an optical frontend do to maximize inference accuracy? When does allocating more of the computation to the optical layer help, and when is it ineffective?
How should the optical structure be designed to achieve high-performance inference? More specifically, do recent advances in meta-optics, such as ``nonlocal'' metasurfaces~\cite{monticone2025nonlocality,overvig2022diffractive}, provide an advantage in this context? Here, we address these questions focusing on \emph{linear} optical frontends, considering both coherent and incoherent illuminations, different forms of nonlocality, and different classes of frontend structures. We use a deliberately simple image-classification task chosen to reveal the underlying tradeoffs cleanly. While other recent works have addressed related questions for end-to-end-optimized systems for \emph{imaging} and image reconstruction~\cite{Pinkard:2024,Kabuli:2026,Miller:2026}, here we specifically consider hybrid optical systems for \emph{inference} and highlight important distinctions between the two cases. In particular, we show that, while intensity detection is a limitation for image reconstruction, it can become a computational resource for classification.

Our results show that a well-designed linear frontend performs \emph{statistical-distribution reshaping}: it increases the class separability of the sensor readout, and this separability, quantified by the Bhattacharyya distance~\cite{Bhattacharyya1943,sethna2006statistical}, is a training-free predictor of downstream accuracy. We validate this correspondence well beyond the regime in which related results can be rigorously proven for incoherent local frontends~\cite{Wang:2026}. Our results further show that much of the discriminative information is encoded in the inter-pixel correlations of the joint readout, and confirm that the optical advantage is restricted to the bottleneck regime. We also demonstrate that, because the sensor records intensity, a linear field operator produces features that are \emph{quadratic} in the input, with the additional class information residing in interference cross-terms that require nonlocal, coherent optics. This allows us to clarify what type of nonlocality is beneficial for inference and to identify an empirical upper bound on the performance of linear optical frontends for the considered inference task.

\section{Results}

\subsection{\label{sec:model}Model: intensity detection makes a linear optical frontend quadratic}

The considered hybrid system (Fig.~\ref{fig:Schematic}) consists of an optimizable optical frontend, a fixed low-resolution intensity sensor, and an optimizable electronic backend.
We consider different types of optical frontends, representing standard phase masks, local and nonlocal metasurfaces, and volumetric metamaterials. We begin with a simple metasurface phase mask. Within scalar diffraction theory, the metasurface imparts a local, position-dependent phase $\phi(x,y)$ across an aperture, described by the transmission function $t(x,y)=A(x,y)\,e^{i\phi(x,y)}$, with $A$ the fixed aperture function. The physical system is a lens-free cascade: the input field propagates a distance $d_1$ from the object plane to the metasurface, is multiplied by $t(x,y)$, and propagates a further distance $z$ to the sensor,
\begin{equation}
U_{\mathrm{out}} \;=\; \mathcal{P}_{z}\!\big[\,t\cdot\mathcal{P}_{d_1}[\,U_{\mathrm{in}}\,]\,\big],
\qquad
\mathcal{P}_{d}=\mathcal{F}^{-1}\,H(k_x,k_y;d)\,\mathcal{F},
\label{eq:cascade}
\end{equation}
where $\mathcal{P}_d$ is the (angular-spectrum) propagation operator in a transversely invariant region of length $d$, with transfer function $H$, and $\mathcal{F}$ is the lateral Fourier transform. In the case of free-space propagation, the transfer function is simply $H_f(k_x,k_y;d)=\exp(ik_zd)=\exp(id\sqrt{k_0^2-k_x^2-k_y^2})$, with $k_0$ the free-space wavenumber. Equation~\eqref{eq:cascade} is space-variant in general, namely, the system has no single point-spread function. Each object point illuminates the mask with its own laterally shifted (and curved) wavefront and is therefore imprinted with a different portion of the phase profile. As a result, a shifted input does not produce a merely shifted output. Only in certain limits does a single convolution kernel emerge, particularly in the incoherent case, as shown in the Supplementary Information. The trainable phase $\phi(x,y)$ and the two propagation legs together determine the realized operator, which is generally diagonal in neither real space (as the mask alone would be) nor Fourier space (as free propagation would be).

Discretizing the input field into a vector $\uin\in\mathbb{C}^{D}$ (one entry per input sample point) and the sensor-plane field into $\uout$, Eq.~\eqref{eq:cascade} is a \emph{linear} map,
\begin{equation}
\uout = M\,\uin,
\qquad
M = P_{z}\,\mathrm{diag}(t)\,P_{d_1},
\label{eq:operator}
\end{equation}
where $P_d$ is the discrete propagation matrix (diagonal in $k$-space, assuming transverse invariance) and $\mathrm{diag}(t)$ is a matrix representing the mask (diagonal in real space, assuming a perfectly local response).
A general volumetric medium used as an optical frontend need not take this form and is not constrained by these assumptions: its most general action is an unconstrained operator, which we use below as an idealized limit.

The sensor measures intensity, $|\,\cdot\,|^2$, and averages it over each pixel (pooling operation). Before pooling, the intensity at sensor point $j$ is
\begin{equation}
I_j = \big|(M\uin)_j\big|^2
= \underbrace{\sum_k |M_{jk}|^2\,|u_k|^2}_{\text{diagonal (incoherent-like)}}
\;+\!\!\underbrace{\sum_{k\neq l} M_{jk}M_{jl}^{*}\,u_k u_l^{*}}_{\text{interference cross-terms}} .
\label{eq:quadratic}
\end{equation}
From this equation, we clearly see that the readout is \emph{quadratic} in the input \emph{field}. The first sum is a non-negative linear combination of input intensities $|u_k|^2$: on its own it is a linear map of the input intensity. The second sum, that is, the interference cross-terms, is what makes a linear optical frontend more than a linear intensity map. This feature has two necessary prerequisites.

\emph{(a) Nonlocality.} A cross-term $M_{jk}M_{jl}^{*}$ with $k\neq l$ is nonzero only if row $j$ of $M$ couples two distinct input points $k$ and $l$ onto the same sensor point, that is, only if $M$ represents a \emph{nonlocal} operation, one whose output at a given point depends on the input over an extended spatial region. Optical systems are generically nonlocal in this sense (with the exception of local phase masks considered by themselves).
In a standard metasurface frontend, all nonlocality in Eq.~\eqref{eq:cascade} is free-space propagation-induced: the matrices $P_{d_1}$ and $P_z$ spread each input point across the mask and the sensor, coupling distinct points $k\neq l$ onto the same pixel. Fully accessing the cross-terms therefore requires sufficient propagation, or strategies that effectively compress space for light propagation, such as so-called nonlocal spaceplates~\cite{reshef2021optic,guo2020squeeze,chen2021dielectric,pahlevaninezhad2024multi}, which implement the transfer function $H_f(k_x,k_y;d)$ of a free-space volume over a smaller thickness. The Supplementary Information quantitatively shows how performance degrades as the propagation length is reduced.

\emph{(b) Coherence.} The cross-terms also require a deterministic relative phase between $u_k$ and $u_l$. Under spatially incoherent illumination the input field is a zero-mean random process with $\langle u_k u_l^{*}\rangle = \delta_{kl}\,\langle|u_k|^2\rangle$, so the cross-terms average to zero and the expected readout collapses to a purely linear map of the input intensity,
\begin{equation}
 \langle I_j\rangle_{\text{incoh}} = \sum_k |M_{jk}|^2\,\langle |u_k|^2\rangle .
 \label{eq:incoherent}
\end{equation}
In this regime, the most a frontend can do is reweight and remap input intensities linearly.
This is precisely the operating regime of conventional incoherent imaging~\cite{Wang:2026}, and of recent natural-light metasurface vision processors, whose metasurface-mask optics realize exactly such a non-negative intensity map while all nonlinearity is supplied by the electronic backend~\cite{Peng:2026}. We confirm the associated performance degradation directly in Sec.~\ref{sec:ceiling}: only under coherent illumination can a (sufficiently nonlocal) optical frontend exceed the performance of the best trained linear preprocessor of the same dimensionality.

\begin{figure*}[htbp]
    \centering
    \includegraphics[width=0.60\textwidth]{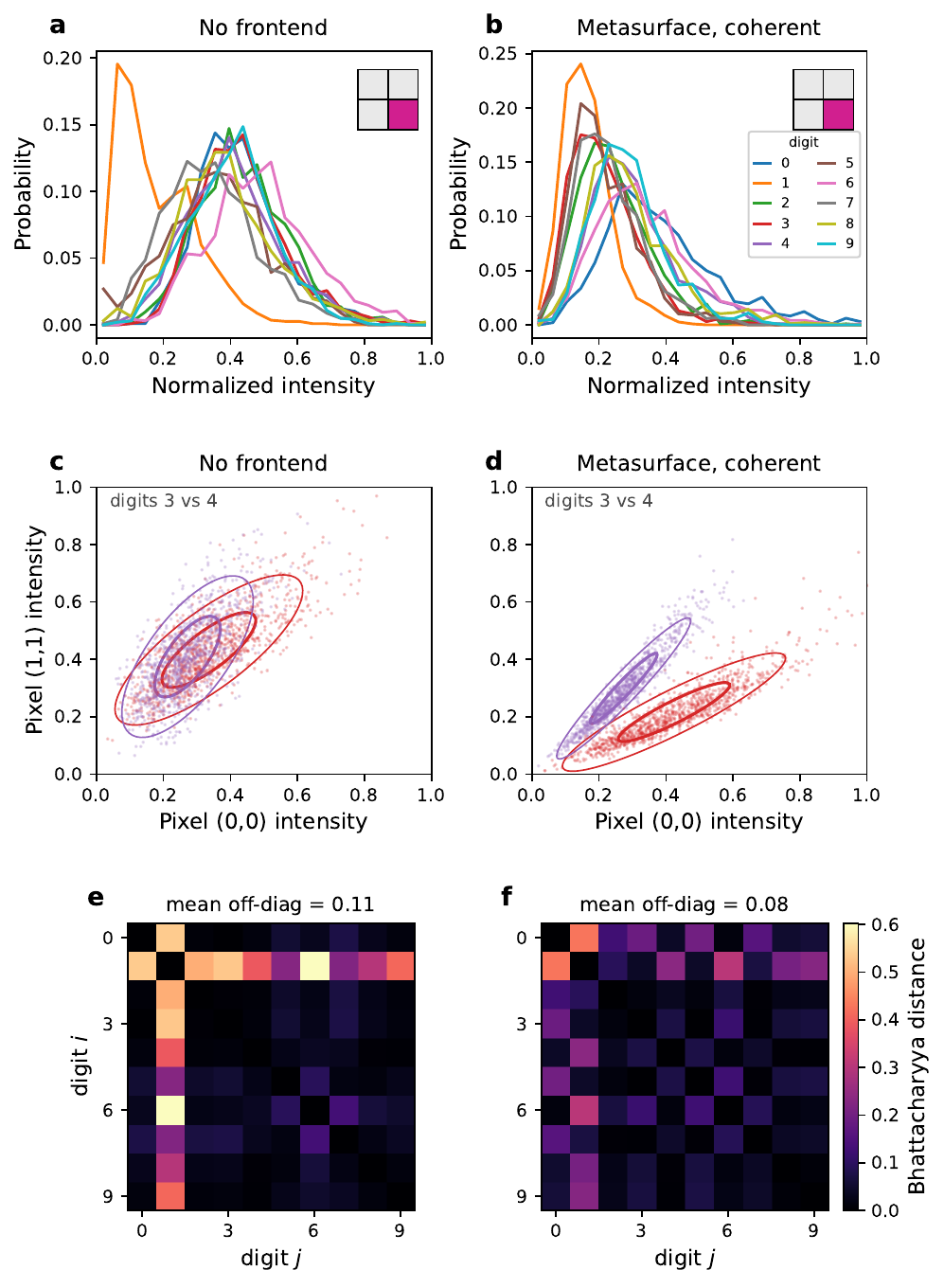}
    \caption{\label{fig:2x2comparison}
    \textbf{Per-pixel versus joint class separability of the $2\times2$-sensor readout.}
    (a,b) Class-conditional intensity-readout distributions at pixel $(1,1)$ (see inset) for digits ``0''--``9'', (a) without a frontend and (b) with the end-to-end-optimized metasurface frontend under coherent illumination: the optimized frontend leaves the single-pixel class distributions as overlapped as in the baseline case.
    (c,d) Joint readout of the pixel pair $\{(0,0),(1,1)\}$ for digits ``3'' (red) and ``4'' (purple), with $1\sigma$ and $2\sigma$ Gaussian ellipses, revealing where the class-discriminative information resides. (c) Without a frontend the two classes overlap. (d) The optimized frontend separates them along a correlated direction (the joint two-pixel Bhattacharyya distance is $2.1$, compared with a best single-pixel value of $0.2$). This separation is invisible in either per-pixel distribution. For illustrative purposes, the pixel and digit pair were selected for maximal correlation gain (the same pair is shown in both panels).
    (e,f) Pairwise Bhattacharyya-distance matrices over all class pairs for the distributions of (a,b), estimated nonparametrically (i.e., without a distributional assumption). The mean off-diagonal separability at this pixel decreases slightly, from $0.11$ to $0.08$, even as downstream test accuracy rises from $0.52$ to $0.80$.
    }
\end{figure*}

\subsection{\label{sec:separability}What the frontend does: statistical separability}

We now discuss and quantify what type of operation a ``well-designed'' optical frontend performs to maximize downstream inference accuracy. 
After the input is preprocessed by the optical frontend, it is measured by a sensor with $N \times N$ pixels. Each pixel records the average intensity over its support, giving a non-negative readout vector $\mathbf{y}=(y_1,\dots,y_{N^2})^\top\in\mathbb{R}^{N^2}_{\ge0}$. Hence, this measurement maps the data into an $N^2$-dimensional readout space. The statistical spread of intensity readouts across the dataset defines a class-conditional \emph{distribution} in this space. Examples of such a distribution across one, two, and three pixels are illustrated in Fig.~\ref{fig:Schematic}(b--d). Such a distribution plays a critical role in the (best possible) performance of the backend classifier. Taking the MNIST classification task as an example, if two classes, e.g., the digits ``1'' and ``7'', are identically distributed in this space, then no backend classifier that takes the readout vector $\mathbf{y}$ as input can distinguish between them. Thus, it is natural to hypothesize that a well-designed optical frontend should reshape these distributions so that the classes are as separable as possible at the bottleneck.

Figure~\ref{fig:2x2comparison} sheds light on this hypothesis, for the case of a $2\times2$ sensor. Pixel by pixel, an end-to-end-optimized \emph{local} metasurface frontend under coherent illumination does not appear to help: its single-pixel class distributions remain as overlapped as those of the no-frontend baseline, as shown in Fig.~\ref{fig:2x2comparison}(a,b). Yet, the same frontend raises downstream test accuracy substantially, from $0.52$ to $0.80$. The reason for this discrepancy is that the optimized frontend encodes the class information into the \emph{inter-pixel correlations} of the joint readout: a pair of classes indistinguishable at every individual pixel can be cleanly separated in the joint distribution of a pixel \emph{pair}, as demonstrated in Fig.~\ref{fig:2x2comparison}(c,d). This correlation-based encoding emerges from conventional end-to-end metasurface optimization, with no separability objective imposed by hand. Physically, this makes sense: a local metasurface phase mask, at a sufficient distance from the sensor (Supplementary Information), is able to redistribute light across the sensor plane---particularly in the coherent case, where interference allows fields coming from different parts of the input to cancel selectively---and, in doing so, reshape the readout distributions to improve class separability.

To make these observations more rigorous, we quantify the separability of two class-conditional readout distributions $p,q$ with the classical Bhattacharyya distance~\cite{Bhattacharyya1943},
\begin{equation}
\DB(p,q) = -\ln\!\int\!\sqrt{p(\mathbf{y})\,q(\mathbf{y})}\;d\mathbf{y}\;\ge 0 .
\label{eq:bhatt}
\end{equation}
We choose this statistical distance for several reasons, as detailed in the Methods. Notably, this metric is directly related to the task under consideration: it upper-bounds the Bayes classification error, $P_{\mathrm{err}}\le\sqrt{\pi_{1}\pi_{2}}\,e^{-\DB}$ for priors $\pi_{1},\pi_{2}$, so a larger $\DB$ tightens the bound on the achievable error~\cite{Kailath:1967}. We note that, in the two-class, shared-covariance Gaussian limit, the Bhattacharyya distance reduces to one eighth of the squared Mahalanobis distance, corresponding to the special case recently studied in~\cite{Wang:2026} for incoherent local frontends.

Applying this metric, per pixel, to Fig.~\ref{fig:2x2comparison}(a,b) quantifies the observations above: without a frontend only class ``1'' stands out (Fig.~\ref{fig:2x2comparison}(e)), and the optimized frontend leaves the per-pixel separability at nearly the same level (Fig.~\ref{fig:2x2comparison}(f)).
Indeed, as discussed above, the discriminative information a frontend imparts is carried by the \emph{joint} statistics of all $N^{2}$ pixels, and a per-pixel measure is blind to the inter-pixel correlations. This is confirmed by evaluating the joint separability using the Bhattacharyya distance for multivariate distributions (Methods).
For the coherent local-metasurface frontend, although the mean per-pixel separability remains at the baseline level, the joint separability is more than twice that of the baseline ($1.64$ versus $0.76$). This confirms that the frontend's $28\%$ inference-accuracy gain is carried by correlations (as further validated in the Supplementary Information),
showing that any proxy metric intended to reliably predict inference-accuracy gains must be correlation-aware and use the joint readout.

\begin{figure*}[htbp]
    \centering
    \includegraphics[width=0.86\textwidth]{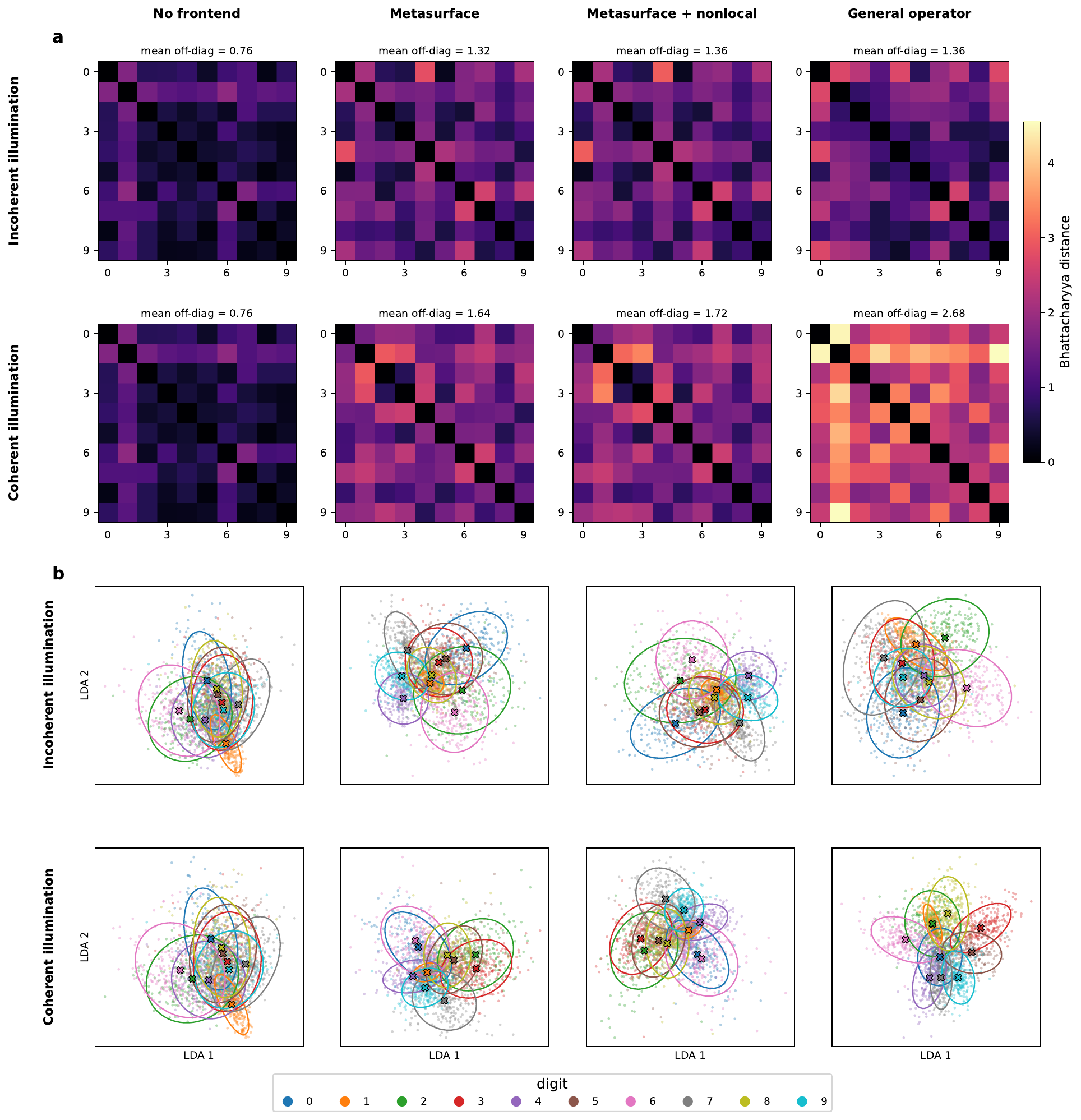}
    \caption{\label{fig:separability}
    \textbf{Joint (correlation-aware) class separability of the 2-by-2-sensor readout} for every configuration, under incoherent (top) and coherent (bottom) illumination, with the frontend increasing in generality from left to right (no frontend; local metasurface; local metasurface + nonlocal $k$-space filter; general operator). The coherent nonlocal-filter column uses the sensor-side placement (the object-side placement yields the same performance; see Supplementary Table~1).
    (a) Pairwise class-conditional Bhattacharyya-distance matrices from a multivariate Gaussian model (the top label indicates the mean off-diagonal value). These results show that separability tracks accuracy: it increases with coherence (top$\to$bottom) and with operator generality (left$\to$right), and the nonlocal amplitude filter does not improve performance over the local metasurface.
    (b) Two-dimensional LDA projection of the same readouts, showing the corresponding class clusters (points). Crosses mark the class means, and ellipses indicate the per-class $2\sigma$ contours. Axes are the first two linear-discriminant components, with their scale arbitrary.
    }
\end{figure*}

Figure~\ref{fig:separability} shows the joint separability for every configuration considered in this paper and both illuminations (the third and fourth columns will be discussed in the next section). Comparing Fig.~\ref{fig:separability}(a) with Fig.~\ref{fig:accuracy}(a,b) for the $2\times2$ case shows that this proxy metric rank-orders the frontends consistently with their measured accuracy (with a Spearman's rank correlation coefficient of $\rho=0.97$ across the eight $2\times2$ configurations; see Methods). These results empirically confirm and extend the separability--accuracy correspondence to ten
classes and full per-class covariances (allowing for class-dependent distributional shapes), beyond the two-class, shared-covariance case in which it was proven for incoherent frontends~\cite{Wang:2026}. Coherence brightens the separability matrices as seen by comparing the top and bottom rows of Fig.~\ref{fig:separability}(a), while operator generality brightens them from left to right. The same ordering is visible directly as increasingly separated class clusters in the distribution projections of Fig.~\ref{fig:separability}(b).

To shed light on the algorithmic functionality of the optical frontend, we specifically examine what type of dimensionality reduction it performs. We can compare its operation with two common linear dimensionality-reduction techniques: linear discriminant analysis (LDA), which maximizes inter-class variance relative to intra-class variance, and principal component analysis (PCA), which maximizes total variance irrespective of class~\cite{Hastie:2009}. Our findings reveal that the optimized optical frontends go clearly beyond PCA and slightly beyond LDA in terms of the statistical structure they create at the readout. The joint separability given by the multivariate Bhattacharyya distance splits into two contributions (Methods):
one from the separation of the class-distribution \emph{means}, namely the centers of the distributions, corresponding to the quantity optimized by LDA for class separation; and one from differences in the class \emph{covariances}, namely the distributional shapes, which are not exploited by standard LDA.
For the classification task considered here, we numerically find that the mean term accounts for 72\% of the coherent local-metasurface frontend's joint separability, with the remaining 28\% arising from covariance differences.

Figure~\ref{fig:2x2comparison}(c,d) illustrates the dominant, LDA-like contribution: the optical frontend transforms two overlapping class distributions into two distributions of similar shape whose means are separated along an oblique direction defined by inter-pixel correlations. The smaller covariance-difference contribution is analogous to the additional statistical information exploited by quadratic discriminant analysis (QDA):
while standard LDA assumes the same covariance for all classes, QDA allows each class to have a distinct covariance and can therefore exploit differences in distributional shape~\cite{Hastie:2009}. In brief, the representation created by the optical frontend is predominantly LDA-like but contains a smaller, genuinely beyond-LDA component. This also connects naturally to the quadratic features of Eq.~\eqref{eq:quadratic}: coherent nonlocal field mixing expands the feature space beyond that available to a linear intensity map, providing an additional mechanism available to the frontend to reshape not only the class means but also the class-dependent covariance structure of the readout (verified in the Supplementary Information).

\subsection{\label{sec:bottleneck}When an optical frontend helps: the sensor bottleneck}

\begin{figure*}[htbp]
    \centering
    \includegraphics[width=0.92\textwidth]{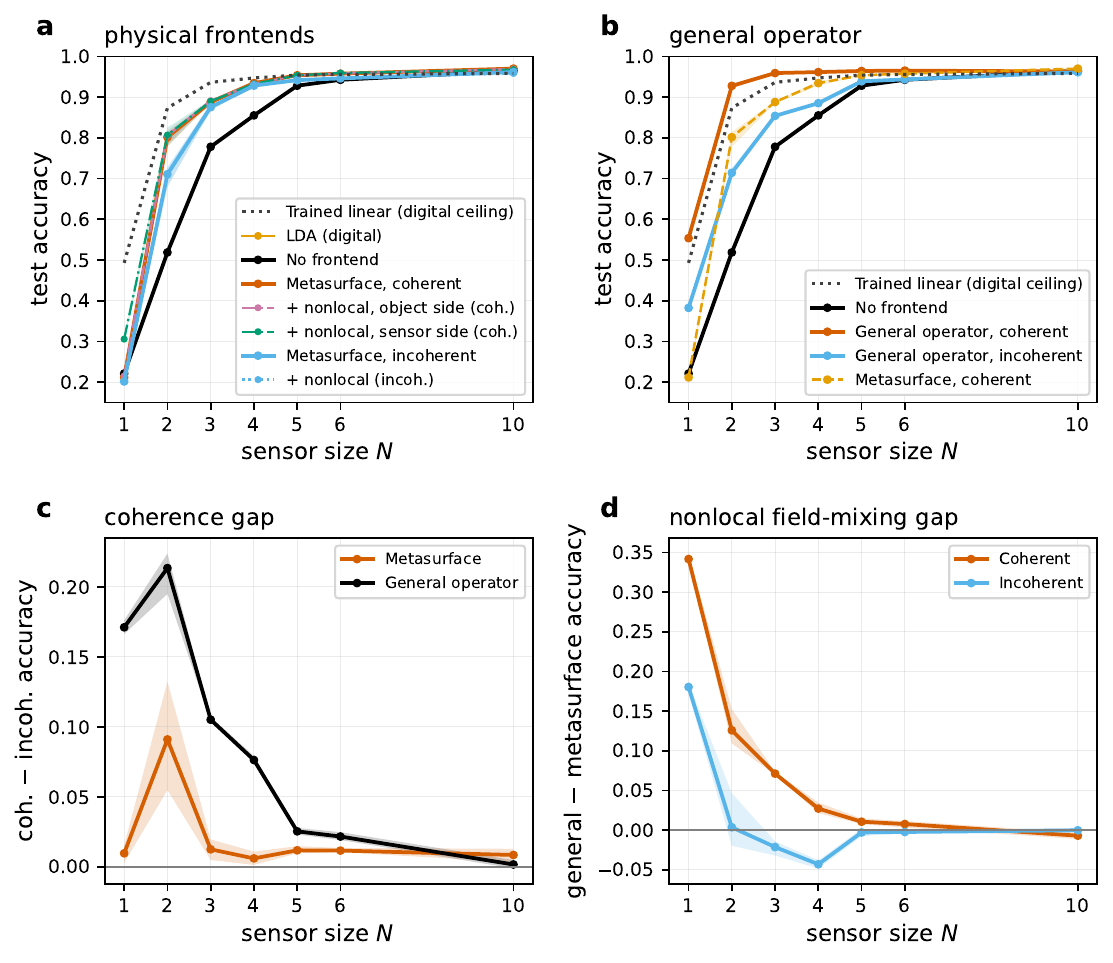}
    \caption{\label{fig:accuracy}
    \textbf{Test accuracy versus sensor size $N\times N$.}
    (a) Physical frontends: a backend-only network (no optimized frontend), the optimized local metasurface, and the metasurface with an added nonlocal amplitude filter on the object side and sensor side, under coherent (solid) and incoherent (dashed) illumination, together with two digital linear references: LDA and a trained linear frontend with signed weights (not realizable by a physical optical frontend, and treated here as an empirical linear upper bound).
    LDA terminates at $3\times3$ ($C-1=9$ components for $C=10$ classes).
    (b) General unconstrained device operator (coherent and incoherent) compared against the same references. Under coherent illumination, its accuracy defines an empirical upper bound on the performance of any linear optical frontend for the considered inference task. Under incoherent illumination, its performance collapses to the metasurface level.
    (c) Coherence accuracy gap (coherent $-$ incoherent accuracy) for the local metasurface and the general device operator.
    (d) Nonlocal field-mixing accuracy gap (general device operator $-$ local metasurface).
    All curves are averaged over three seeds; shaded bands span the seed minimum--maximum range.
    }
\end{figure*}

Having characterized separability, we then ask how performance improves with sensor size, and why. As we have established, the optimized frontend reshapes the distributions of the ten classes so that the joint readout separates them effectively. Specifically, classes indistinct at every individual pixel are separated by the correlations among pixels (Fig.~\ref{fig:2x2comparison}(c,d)), and a sufficiently expressive neural-network backend can then exploit the separation present in any direction of the readout space~\cite{Hornik1989,Ruck1990}. As a result, in the small-sensor regime the hybrid system with an optimized frontend consistently outperforms a backend-only system, as the accuracy-versus-size curves in Fig.~\ref{fig:accuracy}(a) confirm.

As $N$ grows beyond the small-sensor regime, the sensor eventually captures essentially all of the dataset's inherent separability, regardless of the frontend, so the frontend's ability to further separate classes saturates and the backend becomes the limiting factor. We clearly observe this saturation for $N\gtrsim6$ in Fig.~\ref{fig:accuracy}. This result is physically sensible: an optical frontend cannot add task-relevant information as it processes the input signal \cite{Cover2005,Wang:2026}; it can only re-encode the input so that the existing information survives the bottleneck more effectively and can be better exploited by the backend.
Thus, an optimized optical frontend is expected to improve inference accuracy only when the sensor imposes a strong bottleneck on the information flow. We return to this sensor-bottleneck effect, and to its role in determining the optical advantage of coherence and nonlocality, in Sec.~\ref{sec:general}.

\subsection{\label{sec:ceiling}Performance limits and the role of nonlocality, coherence, and quadratic features}
In this section, we address two central questions for linear optical frontends in hybrid inference systems: what is the maximum performance improvement achievable for the considered task, and what physical resources are required to approach this upper bound?

\subsubsection{\label{sec:edge} Nonlocal $k$-space amplitude filtering does not improve performance}
A nonlocal optical element that is transverse-shift-invariant---such as a multilayer slab or a metasurface with subwavelength period, for which the transverse wavevector $\mathbf{k}$ is conserved---acts as a $k$-space transfer function $H(\mathbf{k})$~\cite{Shastri2023,overvig2022diffractive,monticone2025nonlocality}. Modulating its amplitude $|H(\mathbf{k})|$ filters spatial-frequency components: a high-pass filter performs edge or corner detection, a low-pass filter blurs the input image, and a band-pass filter selects features with a specific spatial periodicity. It has remained an open question whether such operations help inference tasks, such as image classification. Intuitively, for example, sharpening edges might help the backend focus on the most important features and thereby classify the input with higher accuracy.

Our analysis provides a quantitative way to test this intuition. A wavevector-conserving $k$-space filter commutes with free-space propagation (both are diagonal in $k$), so its position within a propagation leg is irrelevant. In the cascade geometry of Eq.~\eqref{eq:cascade} there are only two physically distinct placements for such an element, on the object side of the mask or on the sensor side. The two placements perform different operations. On the object side, the filter acts on the input's own spectrum, which free-space propagation leaves unchanged in magnitude (only evanescently attenuating sub-diffractive features), so it is precisely the standard ``optical edge detection'' operation, applied to the field. On the sensor side, it ``prunes'' the propagation channels of the mask-scattered field.

Our results show that neither placement improves performance. An end-to-end-optimized nonlocal amplitude filter added to the local metasurface frontend, in either position, yields accuracy virtually identical to the plain metasurface frontend at every sensor size $N\ge2$ and under either illumination (Fig.~\ref{fig:accuracy}(a)), and nearly identical joint separability (Fig.~\ref{fig:separability}(a)), which rules out a benefit hidden in inter-pixel correlations. A physical interpretation consistent with these results can be seen from Eq.~\eqref{eq:quadratic}: amplitude filtering with $|H|<1$ removes task-relevant photons and, therefore, placed on the object side, it attenuates the $|u_k|^2$ that carry the signal, rather than mixing them, before the mask can exploit them; placed on the sensor side, it attenuates the channels that carry the encoded field to the sensor, reducing the weights $|M_{jk}|^2$ and, with them, the interference cross-terms. In either placement, the added $k$-space flexibility is offset by a reduction in the amount of task-relevant information available to the digital backend. (The only exception is the $1\times1$ sensor case, where the readout is one number, that is, the total transmitted power; thus, a trainable $k$-selective loss channel markedly improves performance by discarding photons class-selectively.) 

Under incoherent illumination the situation is even clearer: the intensity-domain optical transfer function is the \emph{autocorrelation} of its field-domain transmission profile, which necessarily peaks at zero frequency, as shown in~\cite{WangGuo:2020}. Thus, the only intensity filters admitted by incoherent light are low-pass-like or, more precisely, peaked at zero spatial frequency. This operation, however, is similar to the averaging already performed by the sensor. The subtractive, edge-extracting operation that a $k$-space filter performs on a coherent field (through the sign-changing lobes of its real-space kernel) has no incoherent counterpart, because intensity kernels are non-negative. (Incoherent differentiation is achievable only by hybrid optoelectronic schemes that subtract two intensity channels digitally~\cite{WangGuo:2020,Swartz:2024}, moving the signed operation into the electronic backend and doubling the readout).

\subsubsection{\label{sec:general}Coherence and nonlocal field-mixing: toward super-linear performance}
As mentioned in Sec.~\ref{sec:model}, a different kind of nonlocality does help: general nonlocal input-output responses that enable field mixing from different spatial points. In the previous sections, this type of nonlocality was provided entirely by free-space propagation, whose only design parameter is the propagation distance. Here, we consider a more general setting to understand the performance limits of a generic linear optical frontend: a volumetric metamaterial placed in front of the sensor (Fig.~\ref{fig:Schematic}(d)), with a fully optimizable response and without distinguishing between local and nonlocal elements. We model this device as a linear, bounded operator, with no requirement of shift-invariance, symmetry, or Hermiticity. Assuming this is a physical system with a square-integrable Green-function kernel (and spatially separated input and output), the operator is Hilbert--Schmidt and compact, and it can therefore be approximated to any accuracy by a finite matrix $M$~\cite{Miller:2019,miller2026fundamental}, modeling the optical frontend as in Eq.~\eqref{eq:quadratic} (generic linear field transformation followed by intensity detection). This configuration, in which all optical degrees of freedom are fully and independently trainable, achieves a substantially higher accuracy under coherent illumination (Fig.~\ref{fig:accuracy}(b)), revealing a distinct super-linear advantage, as discussed next.

Figure~\ref{fig:accuracy} and Supplementary Table~1 compare the inference accuracy of all frontend configurations considered in this work under both coherent and incoherent illumination, revealing a synergy between two distinct physical resources: \emph{coherence} and \emph{device-operator generality}. As a useful point of comparison, considering the 2-by-2-sensor case, we see that LDA preprocessing of the same dimensionality reaches $80\%$. However, this is not representative of the actual upper bound on what linear preprocessing can achieve for this inference task, as LDA optimizes a specific variance criterion rather than the task itself. By contrast, a general linear frontend of matched dimensionality trained end-to-end with the same backend (see Methods) reaches $87\%$, which we adopt as the empirical linear upper limit/ceiling. Like LDA and PCA, this limit is a digital reference, not a physical device, since it requires \emph{signed} weights, whereas the readout of any optical frontend measured in intensity is non-negative.
(We note, however, that interferometric readout supplying a coherent reference at the sensor would record signed, field-linear terms and could realize signed preprocessing optically, albeit only under coherent illumination, since the reference cross-term would average out for incoherent light.) Around this linear ceiling the measured accuracies organize clearly according to the available physical resources. Incoherent optics fall well below the ceiling: $71\%$ for the metasurface frontend and, interestingly, the same $71\%$ for the unconstrained operator, indicating that, without coherence, operator generality is irrelevant, and the $16\%$ reduction from the linear ceiling is the cost of non-negativity. By contrast, coherent optics allow accessing the quadratic interference cross-terms in Eq.~\eqref{eq:quadratic}, leading to better performance. Coherence increases the metasurface-frontend accuracy by $9\%$, to $80\%$, reaching the performance of LDA. Finally, unconstrained nonlocal field mixing (the general device operator) raises the performance to an impressive $93\%$ under coherent illumination, the only configuration that exceeds the linear ceiling (by $5$--$6\%$), demonstrating genuine super-linear (quadratic) performance, as anticipated in the previous sections.
We also note that Fig.~\ref{fig:accuracy}(c,d) show that, as expected, the advantages provided by coherence and nonlocal field-mixing are concentrated at the tightest bottlenecks ($N\le2$) and provide maximum performance gain only when combined. (Under incoherent illumination, the field-mixing gap even turns negative at moderate sensor sizes: with the cross-terms averaged out, the unconstrained operator adds no expressivity, and its ${\sim}6\times10^{5}$ complex parameters only burden optimization and generalization.)

We treat the accuracy attained by the general device operator under coherent illumination as an \emph{empirical} (and possibly loose) upper bound for optical preprocessing on the considered inference task. Our results indicate that the nonlocality provided by free-space propagation combined with an optimized phase mask is far from sufficient to reach this ``quadratic ceiling,'' because its nonlocal control is weak: this configuration cannot route arbitrary, widely separated input points onto a given pixel with independently controlled magnitudes and phases. Such control is available only to a more general operator, potentially realized by a volumetric metamaterial device. Interesting open questions remain regarding which specific physical properties such a device would need in order to approach this upper limit. These questions will be the subject of future work.

\subsubsection{\label{sec:coherence}The contrast with imaging and communication}
We emphasize again that super-linear performance is a coherent phenomenon. As Eq.~\eqref{eq:incoherent} shows, under incoherent illumination the cross-terms average away, and the readout reduces to a non-negative linear map of input intensity, whose best achievable separability lies below that of the best linear map. Reaching the super-linear ceiling requires \emph{both} coherence and a general nonlocal operator or, in other words, strong field-mixing nonlocality exploited coherently.

These observations sharpen the contrast with imaging and image reconstruction. A recent study on the information limits of imaging under intensity detection~\cite{Miller:2026} showed that the optimal optic is a permutation: each source should be focused onto a distinct detector (``generalized focusing'').
For inference, the conclusion is different. The collapse of complex amplitudes to non-negative intensities, that is, the ``intensity bottleneck'' that limits scene \emph{reconstruction} and makes the optimum a non-mixing, focusing optic~\cite{Miller:2026}, is, for \emph{inference}, precisely the nonlinearity that provides access to quadratic features in the coherent nonlocal case. The same $|\cdot|^2$ operation is therefore a limitation for imaging and a resource for classification.

Interestingly, a similar dichotomy was recently identified at the level of full-wave mutual information in the context of optical information-transfer problems (communication, phase retrieval, and imaging): point focusing was found to be optimal for incoherent sources with uncorrelated statistics, while for coherent sources measured in intensity the optimal operation becomes interferometric mixing under certain conditions, its cross-terms making relative phases information-bearing~\cite{Chen:2026}. This is analogous to the optimal operation we identify for inference, but the operating regimes are opposite: the analysis in Ref.~\cite{Chen:2026} maximizes the information a measurement retains about the \emph{entire source field} (task-free, with no digital backend) and its mixing advantage requires the detectors to \emph{outnumber} the source degrees of freedom. In our setting, the target is a low-dimensional (ten-class) label, not the full field, and mixing becomes beneficial in the opposite, detector-starved regime, where the optics must compress class information through the readout bottleneck (consistent with the information bottleneck principle) rather than recover field information.

\section{Discussion}
In this work, we have shed light on when and how a linear optical frontend can improve performance in hybrid end-to-end-optimized optical--electronic inference systems. We have shown that the main role of a well-optimized optical frontend is to reshape the readout distributions associated with different classes, increasing their joint, correlation-aware separability and thereby easing the task of the digital backend. Another major result is that the largest performance gains require going beyond the incoherent single-metasurface frontend considered in previous works. Because the sensor measures intensity, a linear field operator produces quadratic features (Eq.~\eqref{eq:quadratic}) enabling performance beyond that of any linear-map preprocessor. However, the additional class-discriminative information contained in the interference cross-terms can be exploited only by a frontend that performs sufficiently general nonlocal field mixing under coherent illumination. Our central conclusions are also robust to detection noise over the tested range, as shown in the Supplementary Information.
We also note that these performance gains are concentrated in the sensor-bottleneck regime and saturate as the sensor size increases. More broadly, this diminishing-return behavior echoes our previous finding that performance saturates with increasing device size and structural complexity~\cite{Li:2025}.

Since end-to-end co-optimization of an optical frontend and a digital backend is memory intensive and can face convergence difficulties~\cite{Yang2026}, our separability picture suggests a potentially cheaper and more transparent alternative: optimize the optical frontend alone using a training-free separability objective that promotes joint inter-class separation and intra-class compactness at the sensor plane, and then train the backend to approximate the Bayes-optimal classifier on the resulting representation~\cite{Ruck1990}. The correlation-aware Bhattacharyya separability used here is rank-consistent with downstream accuracy, making it a natural objective for this purpose and, because it bounds the Bayes error, a rational choice. A network with sufficient capacity can then realize the required decision function~\cite{Hornik1989}. Optimizing the optics alone is also cheaper in memory, and a separability-optimized frontend can serve as a good initialization for subsequent end-to-end refinement. Information-theoretic objectives estimated directly from measurements, with no ground truth or decoder in the loop, have recently been shown to design imaging systems that match the performance of end-to-end optimization~\cite{Pinkard:2024,Kabuli:2026}. Our results indicate that a similar approach could also be a natural target for inference frontends. Physically, however, our work reveals a distinct contrast between inference and imaging: while scene reconstruction under intensity detection is optimized by non-mixing, focusing optics~\cite{Miller:2026}, inference is optimized by mixing, nonlocal, coherent optics.

Looking ahead, future work should extend our analysis beyond the scalar-diffraction, single-frequency regime considered here. Polarization- or wavelength-encoded information could provide important additional resources, as already demonstrated in~\cite{Swartz:2024,zheng2024multichannel,WangGuo:2020,zhang2022incoherent}. It will also be interesting to apply our analysis to substantially harder tasks (an initial application to Fashion-MNIST is provided in the Supplementary Information), where we expect our general conclusions to hold but with a wider bottleneck regime and larger performance gaps depending on the dataset. Future work should also identify which physical nonlocal structures can realize, or approach, the general-operator upper bound reported here and identify specific coherent-light scenarios in which the super-linear advantage identified here could have major practical impact. Finally, we believe that the general insights developed here can help guide the design of optical frontends for more complex inference tasks beyond simple classification. Future optical frontends could be trained by \emph{representation learning}~\cite{Bengio2013} to transform raw analog data into embeddings whose task-relevant structure is extracted before digitization and without requiring end-to-end optimization.

We believe that these design principles, combined with the vast parameter spaces offered by volumetric 3D metamaterials and nonlocal meta-optics~\cite{monticone2025nonlocality,overvig2022diffractive,Hu2026,Kalinin2025,Wright2026}, potentially augmented by modest reconfigurability and optical nonlinearity, could substantially expand the capabilities and reach of analog optical preprocessing, paving the way for more efficient hybrid AI systems.

\section{Methods}

\subsection{Dataset and training}
We use MNIST (ten classes): $10{,}000$ training images, of which $1{,}000$ are held out as a validation set, and the standard $10{,}000$ test images. The backend is an MLP with two $256$-neuron hidden layers (ReLU) and a ten-way softmax, trained jointly with the optical-frontend parameters by minimizing the cross-entropy loss (AdamW, batch size $100$, $50$ epochs, exponentially decaying learning rate from $5\times10^{-3}$). Identical split and schedule are used for every configuration. The best epoch is selected on the validation set, and accuracy is reported once on the held-out test set. Every reported accuracy is the mean over three seeds, each seed varying both the train/validation split and the initialization; Fig.~\ref{fig:accuracy} shows the min--max range across seeds as shaded bands.

\subsection{Optical model}
The frontend is simulated using scalar diffraction theory at a free-space wavelength $\lambda=632.8$~nm (HeNe), on a $1200\times1200$ grid with $0.35~\mu$m pitch ($420~\mu$m aperture%
); the input image is upsampled to fill the central $60\%$ of the aperture. In the coherent case, each input grayscale image is encoded in the amplitude of a coherent field with uniform phase. The cascade geometry described by Eq.~\eqref{eq:cascade} is simulated as follows: angular-spectrum propagation over the object--mask distance $d_1=16$~mm, multiplication by the mask, and propagation over the mask--sensor distance $z=4$~mm. This geometry may also include trainable nonlocal elements, applied as
$k$-space amplitude filters within either propagation leg (see ``Nonlocal optics'' below).
The incoherent case uses, instead, an intensity-PSF convolution, which can be derived from the cascade-geometry equation as shown in the Supplementary Information.

\subsection{Local metasurfaces}
The phase profile is parameterized in a Zernike basis (as done, e.g., in Ref.~\cite{Huang2023}), $\phi(x,y)=\sum_{j=1}^{N_Z} w_j Z_j(x,y)$, with trainable coefficients $w_j$ and $N_Z=210$ (Noll indices $1$ to $210$) spanning the full Zernike set (all azimuthal orders $m$) so $\phi(x,y)$ is a smooth phase profile in the 210-dimensional truncated Zernike basis over the aperture.

\subsection{Nonlocal optics}
A transversely invariant, isotropic, nonlocal component can be represented by a $k$-space transfer function $H_A(k_x,k_y)$. 
We consider a trainable amplitude wavevector filter, represented as a smooth annular band-pass with transfer function
\begin{equation}
H_A(k) = \frac{1}{1+e^{-\beta[\kappa-(\kappa_0-\Delta\kappa)]}}
       - \frac{1}{1+e^{-\beta[\kappa-(\kappa_0+\Delta\kappa)]}},
\qquad \kappa = \frac{\sqrt{k_x^2+k_y^2}}{k_0},
\end{equation}
with trainable center $\kappa_0\in[0.2,0.8]$ and full width
$2\Delta\kappa\in[0.4,1.4]$, and fixed edge sharpness $\beta=20$. In the cascade geometry, the filter is placed either on the object side (multiplying $H(k;d_1)$) or on the sensor side (multiplying $H(k;z)$), as discussed in Sec.~\ref{sec:edge}; since the filter commutes with free propagation, its position within a propagation leg is irrelevant.
The band of the $k$-space filter is initialized all-pass, and its width is bounded below ($2\Delta\kappa \ge 0.4$) so that the optimizer can only benefit from, never be trapped by, the filter.

\subsection{Sensor and readout normalization}
The sensor measures intensity ($|\cdot|^2$) and average-pools to one value per pixel. The backend applies batch normalization to the pooled readout. This batch-level normalization is identical across configurations and removes the overall optical scale, so absolute optical gain would play no role in any reported accuracy, in the absence of noise.

\subsection{General device operator}
The idealized general-device frontend is an unconstrained complex matrix $M\in\mathbb{C}^{D\times D}$ applied to the input field, followed by intensity detection (Eq.~\eqref{eq:quadratic}). Its entries are trained end-to-end with no shift-invariance, symmetry, or Hermiticity constraint. Here $D=784$ (corresponding to the $28\times28$ input), so $M\in\mathbb{C}^{784\times784}$ ($\approx1.2\times10^{6}$ real parameters). Selecting the epoch based on validation performance guards against overfitting in this heavily overparameterized operator. While a complex (rather than real) operator is more general and expressive, we found that real and complex operators give statistically indistinguishable accuracy. Because the backend batch-normalizes the readout, the overall scale of $M$ is irrelevant, in the absence of noise: $M$ can be rescaled to $\|M\|_2\le1$ (passive, non-amplifying) without changing the digital readouts or the accuracy. The accuracy it attains is therefore
treated as an empirical upper bound for linear optical preprocessing on this task (whether passive or active, since any linear frontend realizes some matrix in this class).

\subsection{Trained linear frontend}
To directly estimate the performance of the best linear preprocessor, we train an unconstrained linear frontend end-to-end with the same backend: the readout is the pooled output of $W\mathbf{x}$, with $\mathbf{x}$ the vectorized input image and $W\in\mathbb{R}^{784\times784}$ trained jointly with the MLP. Pooling after an unconstrained $W$ spans all linear maps to the readout dimension, so this realizes the empirical linear optimum at each sensor size.
This benchmark is digital, not physical: $W$ has signed entries, whereas the readout of any optical frontend that is measured in
intensity is non-negative (Eqs.~\eqref{eq:quadratic}
and~\eqref{eq:incoherent}). The trained linear frontend exceeds LDA ($87\%$ versus $80\%$ at $2\times2$), as expected since LDA optimizes a specific variance criterion rather than the task loss.

\subsection{\label{app:metrics}Statistical estimation and separability metrics}

Statistical separability is quantified by the Bhattacharyya distance $\DB=-\ln\mathrm{BC}$, with Bhattacharyya coefficient $\mathrm{BC}=\int\!\sqrt{p\,q}$. Per-pixel and joint readout statistics (Figs.~\ref{fig:2x2comparison},~\ref{fig:separability}) are estimated from the full $10{,}000$-image test set ($\sim$1{,}000 per class), exported from the trained models. Readout statistics are computed from the seed-0 models; recomputing on independently retrained models (seeds 1--2) leaves every conclusion unchanged. An important advantage of this statistical distance is that it can be estimated by different procedures and admits a closed-form Gaussian expression in the multivariate Gaussian case.

\emph{Per-pixel separability.} We estimate $\mathrm{BC}$ nonparametrically from the per-class readout histograms (categorical model), making no distributional assumption. In this one-dimensional case the Bhattacharyya distance is monotonically equivalent to the Fisher--Rao distance~\cite{Rao:1945,Miyamoto2024}, the geodesic distance on the statistical manifold under the Fisher information metric: for the categorical model $\dFR^{\mathrm{cat}}=2\arccos\mathrm{BC}\in[0,\pi]$~\cite{Miyamoto2024} and $\DB=-\ln\mathrm{BC}$, both monotonic in $\mathrm{BC}$.
Both the Bhattacharyya and Fisher--Rao distances rank configurations identically.

\emph{Joint (correlation-aware) separability.} A joint histogram estimate over $N^2$ pixels is computationally challenging at realistic sample sizes (a histogram with $B$ bins per pixel has $B^{N^2}$ cells), so we model each class as a multivariate Gaussian and use the closed-form Gaussian Bhattacharyya distance,
\begin{equation}
\DB = \tfrac18(\bm{\mu}_1-\bm{\mu}_2)^{\!\top}\Sigma^{-1}(\bm{\mu}_1-\bm{\mu}_2)
+\tfrac12\ln\frac{\det\Sigma}{\sqrt{\det\Sigma_1\det\Sigma_2}},
\quad \Sigma=\tfrac12(\Sigma_1+\Sigma_2).
\label{eq:gaussbhatt}
\end{equation}
The two terms of this expression separate the two ways in which distributions can differ, as discussed and exploited in Sec.~\ref{sec:separability} to quantify how much of the frontend's separability is of the mean-separation kind. The Fisher--Rao distance, by contrast, has no closed form for general multivariate Gaussians with differing means \emph{and} covariances (only special cases and bounds are known~\cite{Skovgaard1984,CalvoOller1990,Pinele2020}), which is an additional reason to adopt the Bhattacharyya distance for the joint readout.
We emphasize that the Gaussian model is used only as a tractable second-order representation of the joint readout, not a claim that the readouts are Gaussian. Within this Gaussian approximation, the Bhattacharyya distance depends only on the class means and covariances and is invariant under a common translation of the readouts. Thus, the location of the zero-intensity boundary does not enter the calculated distance directly, although the non-Gaussianity associated with non-negative readouts may affect the fidelity of the Gaussian approximation. The adequacy of this approximation is empirically validated in two ways. First, the resulting distances rank-order the frontend configurations consistently with measured accuracy: across the eight $2\times2$ configurations (Fig.~\ref{fig:separability}), the Spearman correlation between the mean pairwise joint distance (three-seed means) and test accuracy is $\rho=0.97$.
Second, at $2\times2$, where the joint space is only four-dimensional, we recomputed all $45$ pairwise distances with a nonparametric $k$-nearest-neighbor estimator of the Bhattacharyya coefficient ($k=10$): the Gaussian and nonparametric distances agree in rank across class pairs (Spearman coefficient of $0.89$--$0.99$ per frontend configuration) and across frontend configurations ($\rho=0.95$), and the nonparametric distances rank-order the configurations against classification accuracy exactly as the Gaussian ones do ($\rho=0.97$).

\subsection{Data availability}
The MNIST and Fashion-MNIST datasets are publicly available. The trained-model readouts and the accuracy data underlying Figs.~\ref{fig:2x2comparison}--\ref{fig:accuracy} and Supplementary Table~1 are available from the corresponding authors on reasonable request.

\subsection{Code availability}
The training, evaluation and analysis code is available from the corresponding authors on reasonable request.

\par\medskip
\noindent\textbf{Contributions}
N.B. performed numerical simulations and analysis. Y.L. and F.M. proposed statistical techniques for the analysis. Y.L. and F.M. prompted Claude Code for further analysis and results presentation. Y.L. and F.M. verified the correctness of the generated code. N.B., Y.L., and F.M. wrote the paper. F.M. and Y.L. supervised the project.

\par\medskip
\noindent\textbf{Competing interests}
The authors declare no competing interests.

\bibliography{nonlocal-optical-neural-net}

\end{document}